\documentstyle[preprint,tighten,aps,floats]{revtex}
\input psfig
\input epsf 

\newcommand{\ifm}[1]{\relax\ifmmode #1\else $#1$\fi}
\newcommand{\etal}{{\it et al.}}
\newcommand{\wb}{\ifm{W}}
\newcommand{\zb}{\ifm{Z}}
\newcommand{\mt}{\ifm{m_T}}
\newcommand{\ppbar}{\ifm{p\overline{p}}}
\newcommand{\Dzero}{D\O}
\newcommand{\dzero}{\Dzero\ Collaboration}
\newcommand{\mw}{\ifm{M_{W}}}
\newcommand{\mz}{\ifm{M_{Z}}}
\newcommand{\qqbar}{\ifm{q\overline{q}}}
\newcommand{\pt}{\ifm{p_T}}
\newcommand{\mpt}{\mbox{$\rlap{\kern0.1em/}\pt$}}
\newcommand{\pe}{\ifm{p(e)}}
\newcommand{\pev}{\ifm{\vec\pe}}
\newcommand{\pte}{\ifm{\pt(e)}}
\newcommand{\ptev}{\ifm{\vec\pte}}
\newcommand{\utv}{\ifm{\vec\ut}}
\newcommand{\ut}{\ifm{u_T}}
\newcommand{\ptnu}{\ifm{\pt(\nu)}}
\newcommand{\ptnuv}{\ifm{\vec\ptnu}}
\newcommand{\ptw}{\ifm{\pt(W)}}
\newcommand{\ptwv}{\ifm{\vec\ptw}}
\newcommand{\wev}{\ifm{W\to e\nu}}
\newcommand{\zee}{\ifm{Z\to ee}}
\newcommand{\wte}{\ifm{W\to \tau\nu\to e\nu\overline\nu\nu}}
\newcommand{\cem}{\ifm{c_{\rm EM}}}

\newcommand{\gt}{\ifm{>}}
\newcommand{\alphaem}{\ifm{\alpha_{\rm EM}}}
\newcommand{\deltaem}{\ifm{\delta_{\rm EM}}}
\newcommand{\alphamb}{\ifm{\alpha_{\rm mb}}}
\newcommand{\srec}{\ifm{s_{\rm rec}}}
\newcommand{\ptee}{\ifm{\pt(ee)}}
\newcommand{\pteev}{\ifm{\vec\ptee}}
\newcommand{\rrec}{\ifm{{\rm R_{rec}}}}

\newcommand{\ueta}{\ifm{u_\eta}}

\newcommand{\pteta}{\ifm{p_\eta (ee)}}
\newcommand{\GEAN}{{\sc geant}}
\newcommand{\alpharec}{\ifm{\alpha_{\rm rec}}}
\newcommand{\betarec}{\ifm{\beta_{\rm rec}}}
\newcommand{\PM}{\ifm{\pm}}
\newcommand{\wth}{\ifm{W\to \tau\nu\to \hbox{hadrons}+X}}
\newcommand{\phie}{\ifm{\phi(e)}}
\newcommand{\PRL}{Phys. Rev. Lett.}
\newcommand{\PL}{Phys. Lett.}
\newcommand{\PR}{Phys. Rev.}
\newcommand{\NP}{Nucl. Phys.}
\newcommand{\NIM}{Nucl. Instrum. Methods in Phys. Res.}
\newcommand{\ZP}{Z.~Phys.}
\newcommand{\uatwo}{UA2 Collaboration}
\newcommand{\cdf}{CDF Collaboration}
\newcommand{\alephc}{ALEPH Collaboration}
\newcommand{\opalc}{OPAL Collaboration}

\begin{document}
\pagestyle{myheadings}
\title{ A Measurement of the \wb\ Boson Mass}
% LIST_OF_AUTHORS.TEX                 12/10/97           
%
\author{                                                                      
%% names begin here                                                           
B.~Abbott,$^{30}$                                                             
M.~Abolins,$^{27}$                                                            
B.S.~Acharya,$^{45}$                                                          
I.~Adam,$^{12}$                                                               
D.L.~Adams,$^{39}$                                                            
M.~Adams,$^{17}$                                                              
S.~Ahn,$^{14}$                                                                
H.~Aihara,$^{23}$                                                             
G.A.~Alves,$^{10}$                                                            
N.~Amos,$^{26}$
% ...
\\                                                                            
\vskip 0.50cm                                                                 
\centerline{(D\O\ Collaboration)}                                             
\vskip 0.50cm                                                                 
}                                                                             
\address{                                                                     
\centerline{$^{1}$Universidad de los Andes, Bogot\'{a}, Colombia}             
\centerline{$^{2}$University of Arizona, Tucson, Arizona 85721}               
\centerline{$^{3}$Boston University, Boston, Massachusetts 02215}             
\centerline{$^{4}$Brookhaven National Laboratory, Upton, New York 11973}      
\centerline{$^{5}$Brown University, Providence, Rhode Island 02912}           
\centerline{$^{6}$Universidad de Buenos Aires, Buenos Aires, Argentina}       
\centerline{$^{7}$University of California, Davis, California 95616}          
\centerline{$^{8}$University of California, Irvine, California 92697}         
\centerline{$^{9}$University of California, Riverside, California 92521}      
\centerline{$^{10}$LAFEX, Centro Brasileiro de Pesquisas F{\'\i}sicas,        
                  Rio de Janeiro, Brazil}                                     
\centerline{$^{11}$CINVESTAV, Mexico City, Mexico}                            
\centerline{$^{12}$Columbia University, New York, New York 10027}             
\centerline{$^{13}$Delhi University, Delhi, India 110007}                     
\centerline{$^{14}$Fermi National Accelerator Laboratory, Batavia,            
                   Illinois 60510}                                            
\centerline{$^{15}$Florida State University, Tallahassee, Florida 32306}      
\centerline{$^{16}$University of Hawaii, Honolulu, Hawaii 96822}              
\centerline{$^{17}$University of Illinois at Chicago, Chicago,                
                   Illinois 60607}                                            
\centerline{$^{18}$Indiana University, Bloomington, Indiana 47405}            
\centerline{$^{19}$Iowa State University, Ames, Iowa 50011}                   
\centerline{$^{20}$Korea University, Seoul, Korea}                            
\centerline{$^{21}$Kyungsung University, Pusan, Korea}                        
\centerline{$^{22}$Institute of Nuclear Physics, Krak\'ow, Poland}            
\centerline{$^{23}$Lawrence Berkeley National Laboratory and University of    
                   California, Berkeley, California 94720}                    
\centerline{$^{24}$Louisiana Tech University, Ruston, Louisiana 71272}        
\centerline{$^{25}$University of Maryland, College Park, Maryland 20742}      
\centerline{$^{26}$University of Michigan, Ann Arbor, Michigan 48109}         
\centerline{$^{27}$Michigan State University, East Lansing, Michigan 48824}   
\centerline{$^{28}$Moscow State University, Moscow, Russia}                   
\centerline{$^{29}$University of Nebraska, Lincoln, Nebraska 68588}           
\centerline{$^{30}$New York University, New York, New York 10003}             
\centerline{$^{31}$Northeastern University, Boston, Massachusetts 02115}      
\centerline{$^{32}$Northern Illinois University, DeKalb, Illinois 60115}      
\centerline{$^{33}$Northwestern University, Evanston, Illinois 60208}         
\centerline{$^{34}$University of Notre Dame, Notre Dame, Indiana 46556}       
\centerline{$^{35}$University of Oklahoma, Norman, Oklahoma 73019}            
\centerline{$^{36}$University of Panjab, Chandigarh 16-00-14, India}          
\centerline{$^{37}$Institute for High Energy Physics, 142-284 Protvino,       
                   Russia}                                                    
\centerline{$^{38}$Purdue University, West Lafayette, Indiana 47907}          
\centerline{$^{39}$Rice University, Houston, Texas 77005}                     
\centerline{$^{40}$Universidade do Estado do Rio de Janeiro, Brazil}          
\centerline{$^{41}$University of Rochester, Rochester, New York 14627}        
\centerline{$^{42}$CEA, DAPNIA/Service de Physique des Particules,            
                   CE-SACLAY, Gif-sur-Yvette, France}                         
\centerline{$^{43}$Seoul National University, Seoul, Korea}                   
\centerline{$^{44}$State University of New York, Stony Brook,                 
                   New York 11794}                                            
\centerline{$^{45}$Tata Institute of Fundamental Research,                    
                   Colaba, Mumbai 400005, India}                              
\centerline{$^{46}$University of Texas, Arlington, Texas 76019}               
\centerline{$^{47}$Texas A\&M University, College Station, Texas 77843}       
}                                                                             
%end                                                                          
\date{\today}
\maketitle
\begin{abstract}
We report a measurement of the \wb\ boson mass based on an integrated luminosity
of 82 pb$^{-1}$ from \ppbar\ collisions at $\sqrt{s}=1.8$~TeV recorded 
in 1994--1995 by the \Dzero\ detector at the Fermilab Tevatron. We identify \wb\
bosons by their decays to $e\nu$
and extract the mass by fitting the transverse mass spectrum from 28{,}323 \wb\
boson candidates. A sample of 3{,}563 dielectron events, mostly due to \zee\
decays, constrains models of \wb\ boson production and the detector. We
measure $\mw=80.44\pm0.10(\hbox{stat})\pm0.07(\hbox{syst})$~GeV. By combining
this measurement with our result from the 1992--1993 data set,
we obtain $\mw=80.43\pm0.11$~GeV. 
\end{abstract}

In the standard model of the electroweak interactions (SM)~\cite{sm}, the mass
of the \wb\ boson is predicted to be 
\begin{equation}
\mw = \left( \frac{\pi\alpha(\mz^2)}
 {\sqrt{2}G_F}\right)^\frac{1}{2} \frac{1}{\sin\theta_w\sqrt{1-\Delta r}}\; .
\label{eq:dr}
\end{equation}
In the ``on-shell'' scheme~\cite{on_shell} $\cos\theta_w = \mw/\mz$, 
where \mz\ is the \zb\ boson mass.  A measurement of \mw,
together with \mz, the Fermi constant ($G_F$), and the electromagnetic 
coupling constant ($\alpha$),
determines the electroweak radiative corrections $\Delta r$ experimentally. 
Purely electromagnetic corrections are absorbed into the value of $\alpha$
by evaluating it at $Q^2=\mz^2$. The dominant contributions to $\Delta r$
arise from loop diagrams that involve the top quark and the Higgs boson. 
If additional particles which couple to the \wb\ boson exist, they will give
rise to additional contributions to $\Delta r$. Therefore, a measurement of \mw\
is one of the most stringent experimental tests of SM predictions. Deviations
from the predictions may indicate the existence of new physics.
Within the SM, measurements of \mw\ and the mass of the top quark constrain  
the mass of the Higgs boson.

This Letter reports a precise new measurement of the \wb\ boson mass based on an
integrated luminosity of 82~pb$^{-1}$ from \ppbar\ collisions at $\sqrt{s}=1.
8$~TeV, recorded by the \Dzero\ detector~\cite{d0_nim} during the 1994--1995 run
of the Fermilab Tevatron. A more complete account of this analysis can be found
in Refs.~\cite{IAdam,EFlattum,prd}. Previously published 
measurements~\cite{UA2,CDF,D0,OPAL,DELPHI,L3,ALEPH}, when combined, determine
the \wb\ boson mass to a precision of 125~MeV.

At the Tevatron, \wb\ bosons are produced mainly through \qqbar\ 
annihilation. We detect them by their decays into electron-neutrino pairs,
characterized by an isolated electron~\cite{ep} with large transverse momentum
(\pt) and significant transverse momentum imbalance (\mpt). 
The \mpt\ is due to the neutrino which escapes detection. 
Many other particles of lower momenta, which
recoil against the \wb\ boson, are produced in the
breakup of the proton and antiproton.
We refer to them collectively as the underlying event.

At the trigger level we require $\mpt>15$~GeV and an energy cluster in the
electromagnetic (EM) calorimeter with $\pt>20$~GeV.
The cluster must be isolated and have a shape consistent with that of an 
electron shower.

During event reconstruction, electrons are identified as energy clusters in the
EM calorimeter which satisfy isolation and shower shape cuts and
have a drift chamber track pointing towards the cluster centroid.
We determine their energies by adding the energy depositions in the first
$\approx40$ radiation lengths of the calorimeter in a window, spanning 0.5 
in azimuth ($\phi$) by 0.5 in pseudorapidity ($\eta$)~\cite{eta}, centered on
the  highest energy deposit in the cluster. Fiducial cuts reject electron
candidates near calorimeter module edges and ensure a uniform calorimeter
response for the selected electrons. The electron momentum (\pev) is 
determined by combining its energy with its direction which is
obtained from the shower centroid position and the drift chamber track.
The trajectories of the electron and the proton beam define the position of the
event vertex. 

We measure the sum of the transverse momenta
of all the particles recoiling against the \wb\ boson, $\utv = \sum_i E_i
\sin\theta_i \hat u_i$, where $E_i$ is the energy deposition in the 
$i^{th}$ calorimeter cell and $\theta_i$ is the angle defined by the cell 
center, the event vertex, and the proton beam. The unit vector $\hat u_i$ points
perpendicularly from the beam to the cell center. 
The calculation of \utv\ excludes the cells occupied by the electron.
The sum of the momentum components along the beam is not well measured because
of particles escaping through the beam pipe. 
From momentum conservation we infer the transverse neutrino momentum, 
$\ptnuv = -\ptev-\utv$, and the transverse momentum of the \wb\ boson,
$\ptwv=-\utv$. 

We select a \wb\ boson sample of 28{,}323 events by requiring 
$\ptnu>25$~GeV, $\ut<15$~GeV, and an electron candidate with 
$|\eta|<1.0$ and $\pte>25$~GeV.

Since we do not measure the longitudinal momentum components of the neutrinos
from \wb\ boson decays, we cannot reconstruct the $e\nu$ invariant mass.
Instead, we extract the \wb\ boson mass from the spectra of the electron \pt\
and the transverse mass, $\mt = \sqrt{2\pte\ptnu(1-\cos\Delta\phi)}$,
where $\Delta\phi$ is the azimuthal separation between the two leptons.
We perform a maximum likelihood fit to the data using probability density
functions from a Monte Carlo program.
Since neither \mt\ nor \pte\ are Lorentz invariants, we have to model
the production dynamics of \wb\ bosons to correctly predict the spectra.
The \mt\ spectrum is insensitive to transverse boosts at leading order in
$\ptw/\mw$ and is therefore less sensitive to the \wb\ boson production model
than the \pte\ spectrum. On the other hand, the \mt\ spectrum depends strongly
on the detector response to the underlying event and is therefore more sensitive
to detector effects than the \pte\ spectrum.

\zb\ bosons decaying to electrons provide an important control
sample. We use them to calibrate the
detector response to the underlying event and to the electrons, and to 
constrain the model for intermediate vector boson production used in the
Monte Carlo simulations.  

A \zee\ event is characterized by two isolated high-\pt\ electrons.
We trigger on events with at least two EM clusters with $\pt\gt 20$~GeV. 
We define two samples of \zee\ decays in this analysis. 
For both \zb\ samples, we require two electron candidates with $\pt>25$~GeV.
For sample~I, we loosen the pseudorapidity cut for one of the electrons to
$|\eta|<2.5$. This selection accepts 2{,}341 events. For sample~II, we require
both electrons with $|\eta|<1.0$ but allow one electron without a matching drift
chamber track. Relaxing the track requirement for 
electrons with $|\eta|<1.0$ increases the efficiency 
%by about 40\% 
without a significant increase in background. 
Sample~II contains 2{,}179 events of which 1{,}225 are in common with sample~I.

For this measurement we developed a fast Monte Carlo program that 
generates \wb\ and \zb\ bosons with the
rapidity and \pt\ spectra given by a calculation using soft gluon 
resummation~\cite{ly} and the MRSA$'$~\cite{mrsa} parton distribution 
functions. 
The line shape is a relativistic Breit-Wigner, skewed by the mass
dependence of the parton luminosity. The measured intrinsic 
widths~\cite{width,ewwg} are used.
The angular distribution of the decay electrons includes a \ptw-dependent 
${\cal O}(\alpha_s)$ correction~\cite{mirkes}. The program also generates
$\wev\gamma$~\cite{berends}, $\zee\gamma$~\cite{berends}, and \wte\ decays.

The program smears the generated \pev\ and \utv\ vectors using a parameterized 
detector response model and applies inefficiencies introduced by the trigger 
and event selection requirements.
The model parameters are adjusted to match the data and are discussed below.

\begin{figure}[p]
\centerline{\psfig{figure=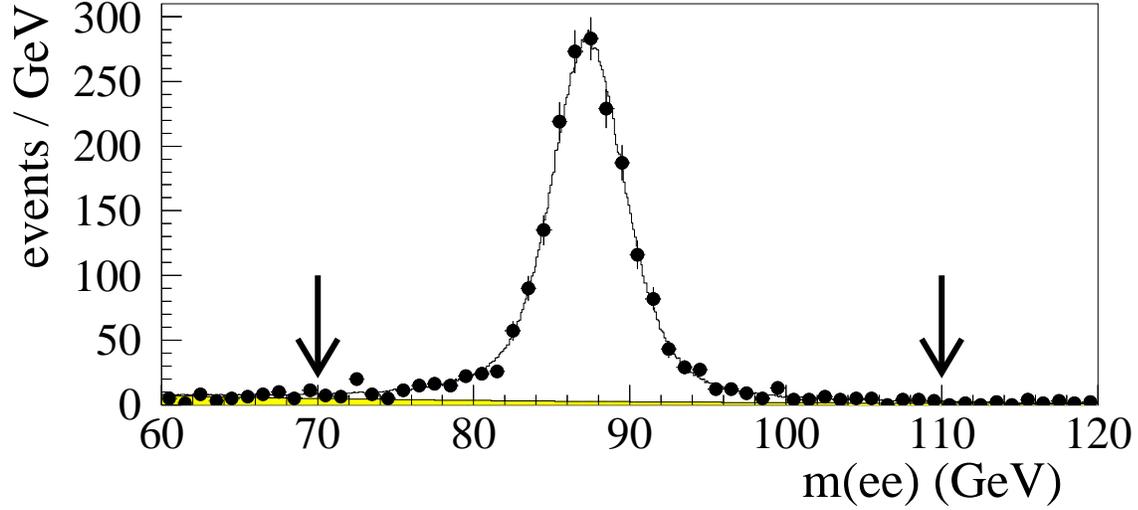,width=7.in,height=7.in}}
\vskip -3in
\caption[]{The dielectron invariant mass distribution of the \zb\ data for 
sample~II ($\bullet$).  The solid line shows the fitted signal plus background
shape and the small shaded area the background. The arrows indicate the fit
window. }
\label{fig:mee}
\end{figure}

The energy resolution for electrons with $|\eta|<1.0$ is described by sampling,
noise, and constant terms. In the Monte Carlo simulation we use a sampling term
of $13\%/\sqrt{\pt/\hbox{GeV}}$, derived from beam tests. The noise term is
determined by pedestal distributions derived from the \wb\ data sample. We
constrain the constant term to $\cem= 1.15^{+0.27}_{-0.36}\%$ by requiring that
the width of the dielectron invariant mass spectrum predicted by the Monte Carlo
simulation is consistent with the \zb\ data (Fig.~\ref{fig:mee}).

Beam tests show that the electron energy response of the calorimeter can be
parameterized by a scale factor \alphaem\ and an offset \deltaem. We determine
these {\it in situ} using $\pi^0\to\gamma\gamma$, $J/\psi\to ee$, and \zee\
decays. 
We obtain $\deltaem = -0.16^{+0.03}_{-0.21}$~GeV and 
$\alphaem =0.9533\pm0.0008$ by fitting the observed mass spectra
while constraining the resonance masses to their 
measured values~\cite{ewwg,pdg}. 
The uncertainty on \alphaem\ is dominated by the finite size of the \zb\ sample.
Figure~\ref{fig:mee} shows the observed dielectron mass spectrum from 
sample~II, and the line shape
predicted by the Monte Carlo simulation for the 
fitted values of \cem, \alphaem, and \deltaem.

\begin{figure}[pth]
\begin{center}
\begin{tabular}{cc}
\epsfxsize = 8.0cm \epsffile[0 0 550 550]{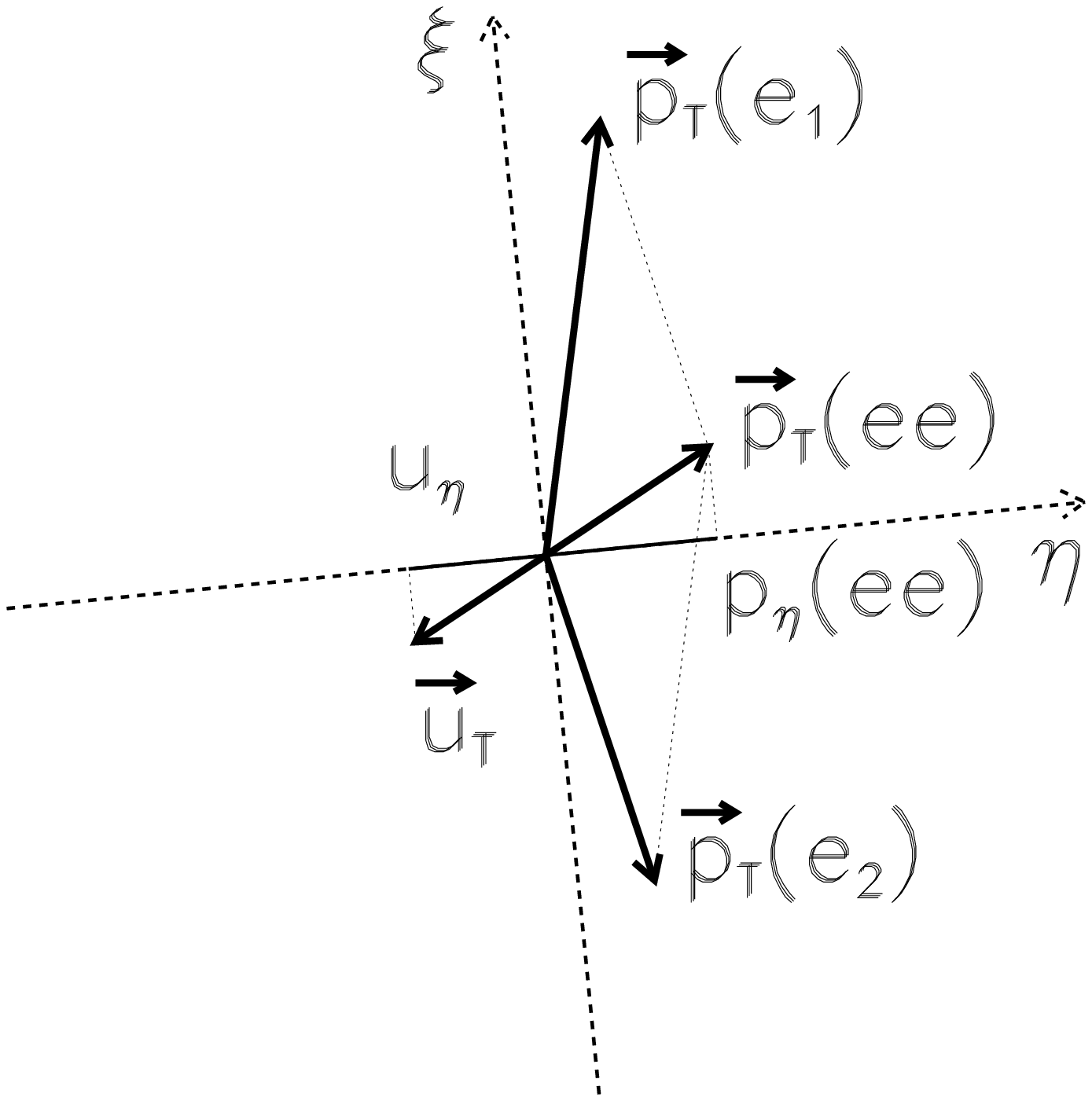} & 
\epsfxsize = 8.0cm \epsffile[0 0 550 550]{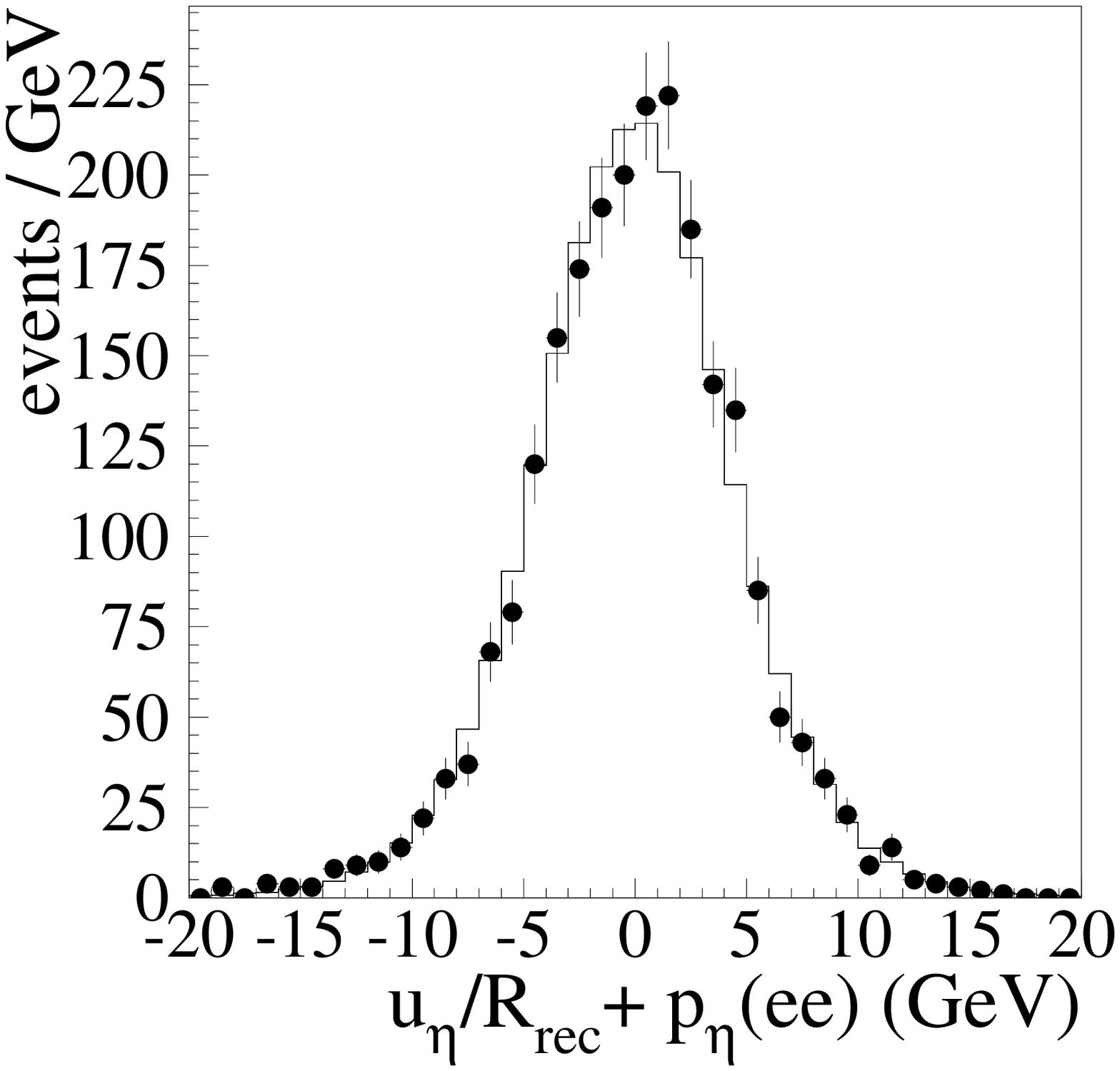} \\ 
\end{tabular}
\caption[]{ The definition of the $\eta$-axis (left).  The plot of
$\ueta/\rrec+\pteta$ (right) for the data ($\bullet$) and simulation~(---).}
\label{fig:hadres}
\end{center}
\end{figure}

We calibrate the response of the detector to the underlying event, relative to
the EM response, using sample~I. The looser rapidity cut
on one electron brings the rapidity distribution of the \zb\ bosons closer to
that of the \wb\ bosons, since there is no rapidity cut on the unobserved
neutrino in \wb\ events. In \zee\ decays, momentum conservation requires  
$\pteev=-\utv$, where \pteev\ is the sum of the two electron \pt\ vectors.
To minimize sensitivity to the electron energy resolution, we project \utv\ and
\pteev\ on the inner bisector of the two electron directions, 
called the $\eta$-axis (Fig.~\ref{fig:hadres}). We call the projections \ueta\
and \pteta.

Detector simulations based on the \GEAN\ program~\cite{geant} 
predict a detector response to the recoil particle momentum of the form
$\rrec = \alpharec + \betarec \ln(\pt/\hbox{GeV})$. We constrain 
\alpharec\ and \betarec\ by comparing the mean value of $\ueta+\pteta$
with Monte Carlo predictions for different
values of the parameters. We measure $\alpharec=0.693\pm0.060$ and 
$\betarec=0.040\pm0.021$ with a correlation coefficient of~$-0.98$.  

The recoil momentum resolution has two components. We smear the magnitude of the
recoil momentum with a resolution of $\srec/\sqrt{\pt/\hbox{GeV}}$.
We describe the detector noise and pile-up, which are independent of the boson
\pt\ and azimuthally symmetric, by adding the \mpt\
from a random \ppbar\ interaction, scaled by a factor \alphamb, to the smeared
boson \pt. To model the luminosity dependence of this resolution
component correctly, the sample of \ppbar\ interactions was chosen to have the
same luminosity spectrum as the \wb\ sample. 
We constrain the parameters by comparing the observed rms of $\ueta/\rrec
+\pteta$ with Monte Carlo predictions and 
measure $\srec=0.49\pm0.14$ and $\alphamb=1.032\pm0.028$ with a
correlation coefficient of $-0.60$.  
Figure~\ref{fig:hadres} shows a plot of $\ueta/\rrec+\pteta$.

%The average transverse energy flow, $S_T = \sum_i
%E_i\sin\theta_i$, excluding the cells occupied by electrons, for the \wb\
%sample is 7.7~GeV higher than for the \zb\ sample.
Excluding the cells occupied by the electrons, the average transverse energy
flow, $S_T = \sum_i E_i\sin\theta_i$, is 7.7~GeV higher for the \wb\ sample than
for the \zb\ sample.  
This bias is caused by requiring the identification of 
two electrons in the \zb\ sample versus one in the \wb\ sample.
The larger energy flow translates into a 
slightly broader recoil momentum resolution in the \wb\ sample. 
We correct \alphamb\ by a factor $1.03\pm0.01$ to
account for this effect in the \wb\ boson model.

Backgrounds in the \wb\ sample are \wte\ decays (1.6\%), 
hadrons misidentified as electrons (1.3\%\PM0.2\%), \zee\ decays 
(0.42\%\PM0.08\%), and \wth\ decays (0.24\%). Their shapes are included 
in the probability density functions used in the fits.

\begin{figure}[htp]
\begin{center}
\begin{tabular}{ll}
\epsfxsize = 8.0cm \epsffile[0 0 550 550]{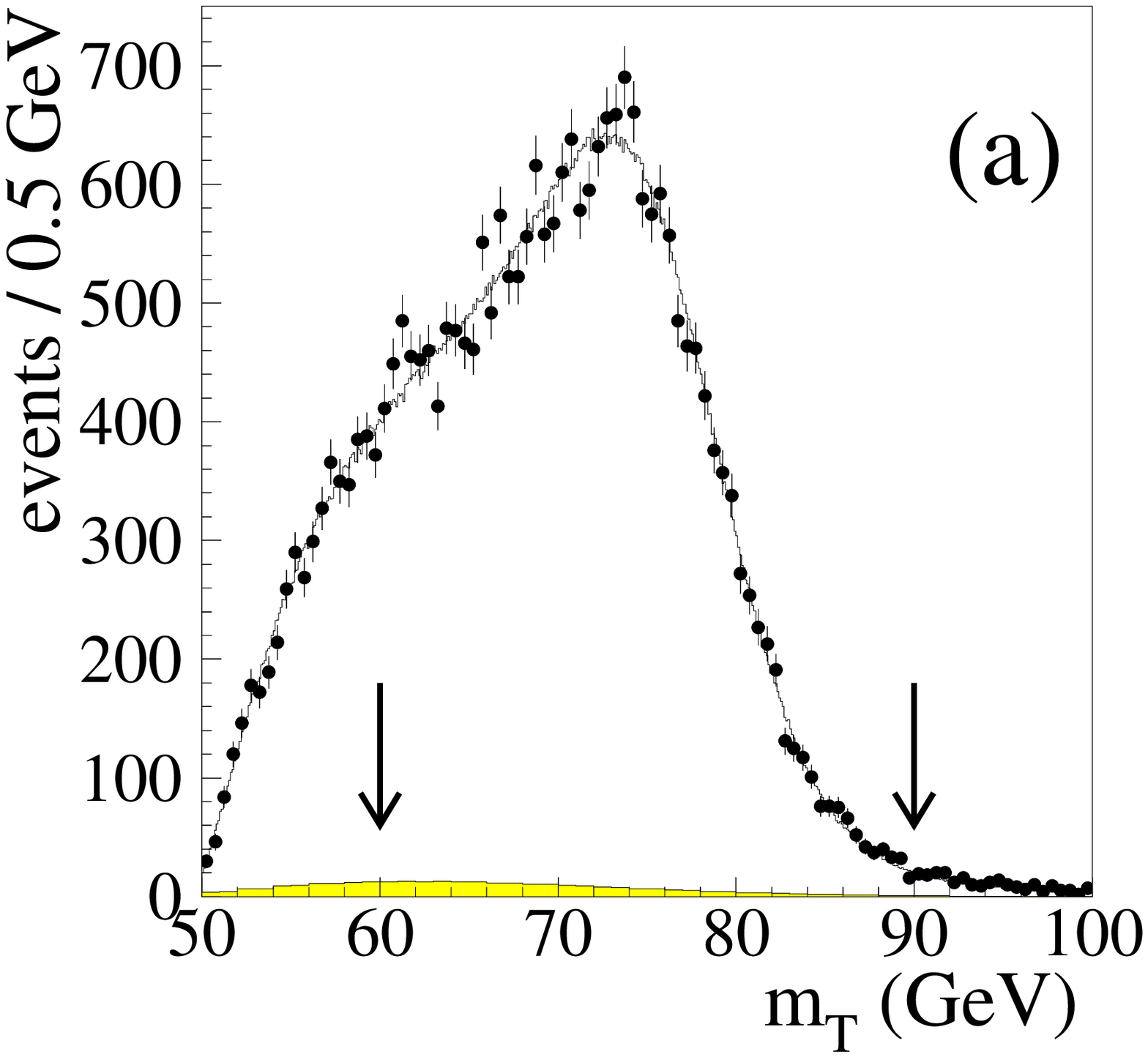} & 
\epsfxsize = 8.0cm \epsffile[0 0 550 550]{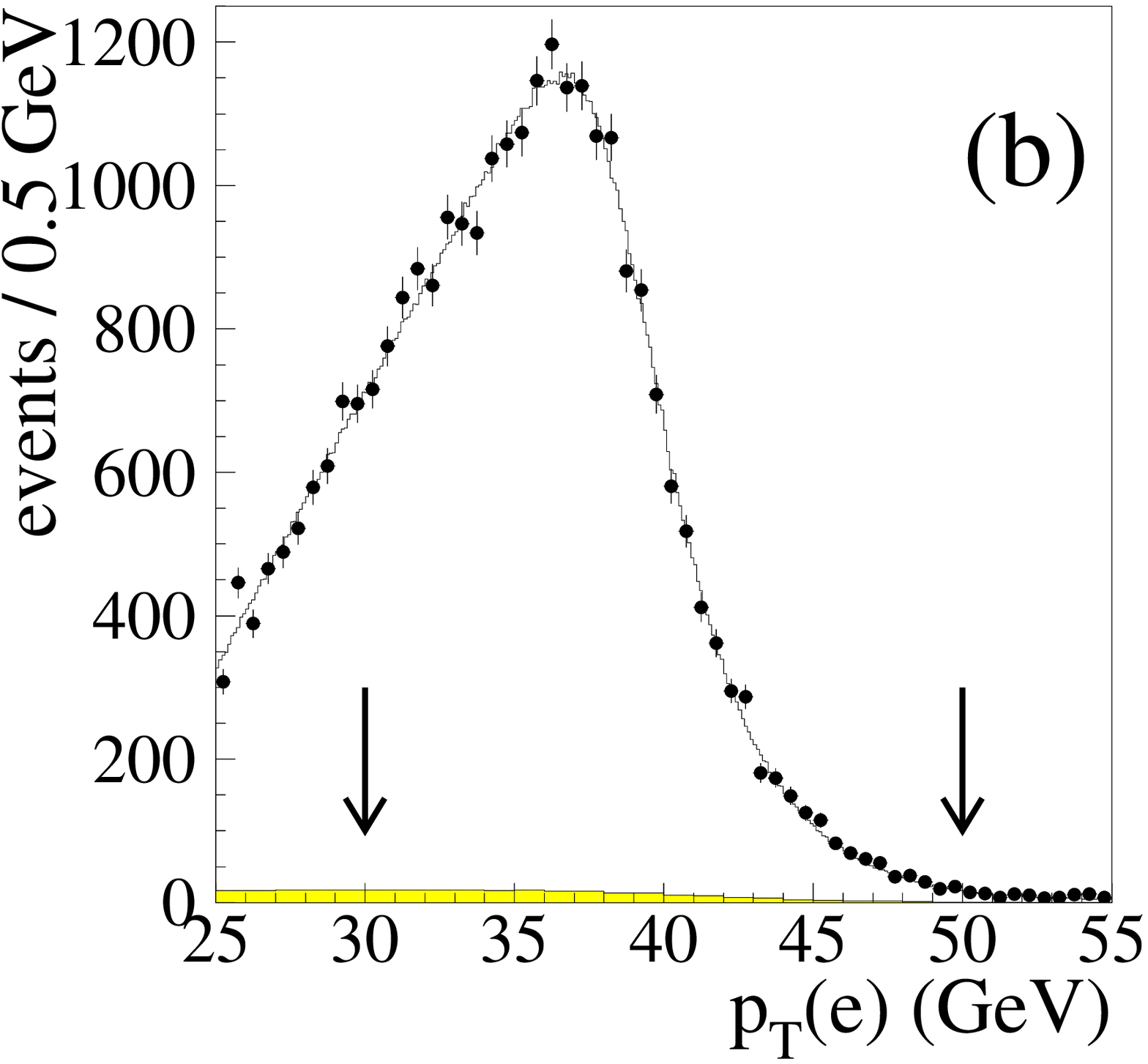} \\ 
\end{tabular}
\caption[]{ Spectra of (a) \mt\ and (b) \pte\ from the data ($\bullet$), 
the fit (---), and the backgrounds (shaded). The arrows
indicate the fit windows.}
\label{fig:mt_fit}
\end{center}
\end{figure}

The fit to the \mt\ distribution (Fig.~\ref{fig:mt_fit}(a)) 
yields $\mw=80.44$~GeV with a statistical uncertainty of 70~MeV. A
Kolmogorov-Smirnov (KS) test gives a confidence level of 28\% that the parent
distribution of the data is the probability density function given by the
Monte Carlo program. A $\chi^2$ test gives $\chi^2=79.5$ for 60 bins which
corresponds to a confidence level of 3\%.
The fit to the \pte\ distribution (Fig.~\ref{fig:mt_fit}(b))
yields $\mw=80.48$~GeV with a statistical uncertainty of 87~MeV.
The confidence level of the KS test is 83\% and that of 
the $\chi^2$ test is 35\%.  

We estimate systematic uncertainties on \mw\ from the Monte
Carlo parameters by varying the parameters within their uncertainties.
Table~I summarizes the uncertainties in the \wb\ boson mass. 
In addition to the parameters described above, the calibration 
of the electron polar angle measurement contributes a significant uncertainty.
We use muons from \ppbar\ collisions and cosmic rays to calibrate the
drift chamber measurements and \zee\ decays to align the calorimeter 
with the drift chambers. Smaller uncertainties are due to
the removal of the cells occupied by the electron from the computation of \utv, 
the uniformity of the calorimeter response, and the 
modeling of trigger and selection biases~\cite{prd}. 

The uncertainty due to the model for \wb\ boson production and decay consists
of several components (Table~I).
%\ref{tab:sum}. 
We assign an uncertainty that
characterizes the range of variations in \mw\ obtained when employing
several recent parton distribution functions: MRSA$'$, 
MRSD$-'$~\cite{mrsd}, CTEQ2M~\cite{cteq2m}, and CTEQ3M~\cite{cteq3m}. We allow 
the \ptw\ spectrum to vary within constraints derived from the \ptee\ spectrum
of the \zb\ data~\cite{prd} and from $\Lambda_{QCD}$~\cite{pdg}. The 
uncertainty due to radiative decays contains an estimate of the effect of
neglecting double photon emission in the Monte Carlo 
simulation~\cite{baur_twophoton}.

The fit to the \mt\ spectrum results in a \wb\ boson mass of 
$80.44\pm0.10(\hbox{stat})\pm0.07(\hbox{syst})\ \hbox{GeV}$ and the fit to
the \pte\ spectrum results in $80.48\pm 0.11(\hbox{stat}) 
\pm0.09(\hbox{syst})\ \hbox{GeV}$. 
The good agreement of the two fits shows that our simulation models 
the \wb\ boson production dynamics and the detector response well.
We have performed additional consistency checks. A fit to the \ptnu\
distribution
yields $\mw=80.37\pm0.12(\hbox{stat})\pm0.13(\hbox{syst})$~GeV, consistent 
with the \mt\ and \pte\ fits. Fits to the data in bins of
luminosity, \phie, $\eta(e)$, and $\ut$ do not show evidence for any 
systematic biases.  

We combine the results from the \mt\ fit and the data collected by \Dzero\ 
in 1992--1993~\cite{D0} to obtain $\mw = 80.43\pm0.11$~GeV.  This is the 
most precise measurement of the \wb\ boson mass to date.
This result is in agreement with the prediction of $80.278\PM0.049$~GeV
from a global fit to electroweak data~\cite{ewwg}.
%from a global fit to the electroweak data from LEP~\cite{ewwg}.
Using Eq.~\ref{eq:dr} we find $\Delta r = -0.0288\pm0.0070$, which 
establishes the existence of electroweak corrections to \mw\ at the 
level of four standard deviations. 

% Acknowledgement_paragraph.tex                             06/18/97
%
We wish to thank U. Baur for helpful discussions.
We thank the staffs at Fermilab and collaborating institutions for their
contributions to this work, and acknowledge support from the 
Department of Energy and National Science Foundation (U.S.A.),  
Commissariat  \` a L'Energie Atomique (France), 
State Committee for Science and Technology and Ministry for Atomic 
   Energy (Russia),
CNPq (Brazil),
Departments of Atomic Energy and Science and Education (India),
Colciencias (Colombia),
CONACyT (Mexico),
Ministry of Education and KOSEF (Korea),
CONICET and UBACyT (Argentina),
and CAPES (Brazil).

\begin{table}[pht]
\begin{center}
\caption{Uncertainties in the \wb\ boson mass measurement in MeV, 
rounded to the nearest 5 MeV.}
\begin{tabular}{lcc}
Source of uncertainty                      & \mt\ fit & \pte\ fit \\ \hline
\ \ \ \wb\ sample size                     & ~70   & ~85   \\
\ \ \ \zb\ sample size (\alphaem)          & ~65   & ~65   \\ \hline
Total statistical uncertainty              & ~95   & 105   \\ \hline\hline
\ \ \ calorimeter linearity (\deltaem)     & ~20   & ~20   \\
\ \ \ calorimeter uniformity               & ~10   & ~10   \\
\ \ \ electron resolution (\cem)           & ~25   & ~15   \\
\ \ \ electron angle calibration           & ~30   & ~30   \\
\ \ \ electron removal                     & ~15   & ~15   \\
\ \ \ selection bias                       & ~~5   & ~10   \\
\ \ \ recoil resolution (\alphamb,\srec)   & ~25   & ~10   \\
\ \ \ recoil response (\alpharec,\betarec) & ~20   & ~15   \\ \hline
Total detector systematics                 & ~60   & ~50   \\ \hline\hline
Backgrounds                                & ~10   & ~20   \\ \hline\hline
\ \ \ \ptw\ spectrum                       & ~10   & ~50   \\ 
\ \ \ parton distribution functions        & ~20   & ~50   \\
\ \ \ parton luminosity                    & ~10   & ~10   \\
\ \ \ radiative decays                     & ~15   & ~15   \\ 
\ \ \ \wb\ boson width                     & ~10   & ~10   \\ \hline
Total \wb\ boson production and decay model & ~30   & ~75   \\ \hline\hline
Total                                      & 115   & 140   \\
\end{tabular}
\end{center}
\end{table}

\begin{references}
\bibitem{ep}
    We generically refer to electrons and positrons as electrons.
\end{references}
\end{document}